\documentclass[12pt]{article}
\newcommand{\be} {\begin{equation}}
\newcommand{\ee} {\end{equation}}
\newcommand{\bdm} {\begin{displaymath}}
\newcommand{\edm} {\end{displaymath}}
\newcommand{\bc} {\begin{center}}
\newcommand{\ec} {\end{center}}
\newcommand{\beqa} {\begin{eqnarray}}
\newcommand{\eeqa} {\end{eqnarray}}

\newcommand{\ra} {\rightarrow}
\begin{document}
\vbox{
\vspace{6.5truemm}}
\bc
{\LARGE\bf Light-Quark Vector-Meson Spectroscopy}
\ec
\medskip
\bc
{\large A. Donnachie$^{*}$ and Yu S Kalashnikova$^{\dagger}$}
\ec
\medskip
\bc
{\footnotesize\it $^*$ Department of Physics and Astronomy, University of 
Manchester,\\ 
Manchester M13 9PL, England}\\
{\footnotesize\it $^\dagger$ ITEP, 117259 Moscow, Russia}
\ec
\medskip
{\footnotesize {\bf Abstract.} The current situation for vector meson 
spectroscopy is outlined, and it is shown that the data are inconsistent 
with the generally accepted model for meson decay. A possible resolution 
in terms of exotic mesons is given. Although this resolves some of the 
issues, fresh theoretical questions are raised.}
\vspace{7truemm}
\bc
{\large\bf INTRODUCTION}
\ec

It is now 15 years since it was first suggested [1,2] that the 
$\rho'(1600)$, as it was then known, is in fact a composite structure,
consisting of at least two states: the $\rho(1450)$ and $\rho(1700)$.
Their existence, and that of their isoscalar counterparts, the 
$\omega(1420)$ and $\omega(1650)$, and of an associated hidden-strangeness 
state, the $\phi(1680)$, is now well established [3]. So it is
pertinent to ask why the light-quark vector mesons remain an important
field of study. The answer is straightforward. Although there is general 
consensus on the existence of these states, there is considerable disparity 
on their masses and widths. Further what is known about the composition of
their hadronic decays raises fundamental questions about the nature of these
states and our understanding of the mechanism of hadronic decays. 

A further complication has been highlighted in two of the talks [4,5] 
at this meeting, with indications of an isovector vector 
meson at a mass of around 1200 to 1250 MeV. This revives an old controversy. 
Many years ago evidence was presented [6] for two vector states with 
masses $1097 \pm 19$ MeV and $1266 \pm 35$ MeV in the reaction 
$\gamma p \to e^+e^- p$. The evidence for these two states was obtained 
from the interference between the Bethe-Heitler amplitude and the real part 
of the hadronic photoproduction amplitude. Additionally in $\omega\pi$ 
photoproduction [7,8,9], $\gamma p \to (\omega\pi) p$, 
the $\omega\pi$ system is dominated by a low-mass enhancement with a peak 
at about 1250 MeV and it seemed natural to associate this with a vector 
state. However it appears that this enhancement is dominated by the 
$J^P = 1^+$ $b_1(1235)$ meson. The evidence for this comes from the analysis
[8,9] of the decay angular distributions of the $\omega\pi$ 
system. The conclusion of [8] is that the data are best described 
by production of the $b_1(1235)$ together with a small $J^P = 1^-$ 
contribution. Additionally the production mechanism does not appear to 
conserve $s$-channel helicity. This latter conclusion is confirmed in 
the experiment of [9] which had the benefit of a linearly-polarised 
beam. The angular distribution of the production plane relative to the photon 
polarisation vector has structure which is inconsistent with $s$-channel 
helicity conservation. The decay angular distributions of the two experiments 
agree and it was also concluded by [9] that the data favour a 
$b_1(1235)$ interpretation over a vector-meson interpretation. Preliminary 
data [10] on $\omega\pi$ photoproduction at high energy indicate a 
cross section for the 1250 MeV enhancement which is comparable to or larger 
than the cross section at much lower energy [8,9], with the natural inference 
that it is being produced diffractively. However diffractive photoproduction 
of the $b_1(1235)$ is inconsistent with all we know about diffraction from 
other reactions. It violates the Gribov-Morrison rule [11,12]
and, more seriously, at the parton level it requires both spin flip and 
angular-momentum flip. The photon is some combination of $^3S_1$ and 
$^3D_1$ $q{\bar q}$ states and the $b_1(1235)$ is $^1P_1$. To avoid conflict 
with diffraction phenomenology, a possible interpretation of the $\omega\pi$
photoproduction data is that it is a mix of $1^-$ and $1^+$ at the lower 
energies, and entirely $1^-$ at high energy, requiring a vector meson at 
about 1250 MeV.

Obviously the light-quark vector mesons present an exciting theoretical 
challenge.

Information on the vector states comes principally from $e^+e^-$ annihilation, 
and also $\tau$ decay for the isovector states, but there are problems with 
much of the data:

$\bullet$ inconsistencies, even in recent high-statistics data

$\bullet$ restricted energy ranges, e.g. Novosibirsk and CLEO

$\bullet$ poor statistics in some channels and missing channels

$\bullet$ inadequate knowledge of multiparticle final states

Fortunately this is set to change with a range of possible new facilities
for $e^+e^-$ annihilation and $\tau$ decay:

$\bullet$ upgrade of Novosibirsk to higher energy

$\bullet$ the PEP-N proposal at SLAC

$\bullet$ the use of initial state radiation (ISR) at BABAR

$\bullet$ emphasis on $\tau$ and charm at CLEO

There is also the possibilty of complementary data on vector meson 
photoproduction:

$\bullet$ from the upgrade of CEBAF

$\bullet$ from real photon radiation in proton-ion and ion-ion collisions 
at RHIC

$\bullet$ photo- and electroproduction at HERA

So the future study of vector mesons looks healthy. 
\vspace{7truemm}
\bc
{\large\bf THE DATA}
\ec

The key data in determining the existence of the two isovector states were 
$e^+e^- \ra \pi^+\pi^-$ [13] and $e^+e^- \ra \omega\pi$ [14].
These original data sets have subsequently been augmented by data on the 
corresponding charged channels in $\tau$ decay [15,16], to which they 
are related by CVC. These new data confirm the earlier conclusions. The 
data on $e^+e^- \ra  \pi^+\pi^-\pi^+\pi^-$ [17] and $e^+e^- \ra 
\pi^+\pi^-\pi^0\pi^0$ [17,18] (excluding $\omega\pi$) and the 
corresponding charged channels in $\tau$ decay [16] are consistent 
with the two-resonance interpretation [19,20], although they do not 
provide such good discrimination. It was also found that the $e^+e^- \ra 
\eta\pi^+\pi^-$ cross section is better fitted with two interfering 
resonances than with a single state [21].  Independent evidence for 
two $J^P = 1^-$ states is provided in a high statistics study of the
$\eta\pi\pi$ system in $\pi^- p$ charge exchange [22]. Decisive 
evidence for both the $\rho(1450)$ and $\rho(1700)$ in their $2\pi$ and
$4\pi$ decays has come from the study of $\bar{p} p$ and $\bar{p} n$ 
annihilation [23]. The data initially available for the study of 
the $\omega(1420)$ and $\omega(1600)$ were $e^+e^- \ra \pi^+\pi^-\pi^0$
(which is dominated by $\rho\pi$) and $e^+e^- \ra \omega\pi^+\pi^-$ 
[24]. The latter cross section shows a clear peak which is apparently
dominated by the $\omega(1600)$. The former cross section is more sensitive
to the $\omega(1420)$. However the only channel in $e^+e^-$ annihilation 
and $\tau$ decay with really consistent data sets over a wide energy range 
is the $\pi\pi$ channel, and that runs out of statistics at the upper end 
of the relevant energy range. 

In addition to the direct experimental problems there are theoretical 
uncertainties which affect the analysis of $e^+e^-$ annihilation and $\tau$ 
decay, and which present data are insufficiently precise to resolve. Firstly 
there is the ``tail-of-the-$\rho$'' problem. In some channels, most notably 
$\pi\pi$ and $\pi\omega$, there is strong interference between the high-energy 
tail of the $\rho$ and the higher-mass resonances. The magnitude and shape
of 
this tail are not known with any precision. They can only be specified in 
models 
and strictly should be part of the parametrisation. Different models yield 
different results for the masses and widths of the resonances. A related 
problem
is the question of the relative phases. These can be specified in simple 
models,
but we know that these models are not precise and leaving the phases as free 
parameters has a major effect on the results of any analysis. 

The experimental challenge is easily stated; high-statistics excitation 
curves for 
a wide range of hadronic final states:
\be
\pi\pi~~~\omega\pi~~~a_1\pi~~~h_1\pi~~~\rho\rho~~~\rho(\pi\pi)_S~~~K{\bar K}
~~~K^*{\bar K}\cdots
\ee
Note that the $n{\bar n}$ states can decay to $K{\bar K}$, $K^*{\bar K}$ etc.
with significant partial widths, so isospin separation is necessary in these 
channels, and there can be mixing between the isoscalar $n{\bar n}$ states 
and the $s{\bar s}$ states.
\vspace{7truemm}
\bc
{\large\bf THE THEORETICAL PROBLEM}
\ec

Despite these various difficulties, an apparently natural explanation for the 
higher-mass vector states is that they are the first radial, $2^3S_1$, and 
first 
orbital, $1^3D_1$, excitations of the $\rho$ and $\omega$ and the first radial 
excitation of the $\phi$, as the generally-accepted masses [3] are close 
to those predicted by the quark model [25]. However this argument is
suspect as the masses of the corresponding $J^P = 1^-$ strange mesons are
less than the predictions, particularly for the $2^3S_1$ at $1414 \pm 15$ MeV
[3] compared to the predicted 1580 MeV [25]. Quite apart from
comparing predicted and observed masses, one would expect the $n{\bar n}$
mesons to be 100 to 150 MeV lighter than their strange counterparts,
putting the $2^3S_1$ at less than 1300 MeV and the $1^3D_1$ below 1600 MeV.
Also this interpretation faces a more
fundamental problem. The data on the $4\pi$ channels in $e^+e^-$ annihilation 
are not compatible with the $^3P_0$ model [26,27,28,29] which is 
accepted as the most successful model of meson decay. The  model works 
well for decays of established ground-state mesons:

$\bullet$ widths predicted to be large, are found to be so

$\bullet$ widths predicted to be small, are found to be so

$\bullet$ calculated widths agree with data to $25 - 40\%$

$\bullet$ signs of amplitudes are correctly predicted

As far as one can ascertain the $^3P_0$ model is reliable, but it has
not been seriously tested for the decays of excited states.

The $^3P_0$ model predicts that the decay of the isovector $2^3S_1$
to $4\pi$ is extremely small:
\be
\Gamma_{2S \to a_1\pi} \sim 3 {\rm MeV}~~~~~~\Gamma_{2S \to h_1\pi} 
\sim 1 {\rm MeV}
\ee
and for the isovector $1^3D_1$ the $a_1\pi$ and $h_1\pi$ decays are large and
equal:
\be
\Gamma_{1D \to a_1\pi} \sim \Gamma_{1D \to h_1\pi} \sim 105 {\rm MeV}
\ee
As $h_1\pi$ contributes only to the $\pi^+\pi^-\pi^0\pi^0$
channel in $e^+e^-$ annihilation, and $a_1\pi$ contributes to both
$\pi^+\pi^-\pi^+\pi^-$ and $\pi^+\pi^-\pi^0\pi^0$, then after subtraction 
of the $\omega\pi$ cross section from the total $\pi^+\pi^-\pi^0\pi^0$
the $^3P_0$ model predicts:
\be
\sigma(e^+e^- \to \pi^+\pi^-\pi^0\pi^0) > \sigma(e^+e^- \to
\pi^+\pi^-\pi^0\pi^0)
\ee
This contradicts observation over most of the available energy range.
Further, and more seriously, it has been shown recently by the CMD 
collaboration at Novosibirsk [30] and by CLEO [31]
that the dominant channel by far in $4\pi$ (excluding $\omega\pi$) up to 
$\sim 1.6$ GeV is $a_1\pi$. This is quite inexplicable in terms of the $^3P_0$ 
model. So the standard picture is wrong for the isovectors, and there are 
serious inconsistencies in the isoscalar channels as well. One possibility 
is that the $^3P_0$ model is simply failing when applied to excited states, 
which is an intriguing question in itself. An alternative is that there is 
new physics involved. 
\vspace{7truemm}
\bc
{\large\bf POSSIBLE SOLUTIONS}
\ec

A favoured hypothesis is to include vector hybrids [32,33], that is 
$q{\bar q}g$ states. The reason for this is that, firstly, hybrid states occur 
naturally in QCD, and secondly, that in the relevant mass range the dominant 
hadronic decay of the isovector vector  hybrid $\rho_H$ is believed to be 
$a_1\pi$ [33]. The masses of light-quark hybrids have been obtained in 
lattice-QCD calculations [34,35,36,37], although with quite 
large errors. Results from lattice QCD and other approaches, such as the bag 
model [38,39], flux-tube models [40], constituent gluon models 
[41] and QCD sum rules [42,43], show considerable variation 
from each other. So the absolute mass scale is somewhat imprecise, predictions 
for the lightest hybrid lying between 1.3 and 1.9 GeV. However it does seem 
generally agreed that the mass ordering is $0^{-+} < 1^{-+} < 1^{--} < 2^{-+}$.

Evidence for the excitation of gluonic degrees of freedom has emerged in
several processes. Two experiments [44,45] have evidence for an 
exotic $J^{PC} = 1^{-+}$ resonance, $\hat\rho(1600)$ in the $\rho^0\pi^-$ 
channel in the reaction $\pi^- N \to (\pi^+\pi^-\pi^-) N$. A peak in the 
$\eta\pi$ mass spectrum at $\sim 1400$ MeV with $J^{PC} = 1^{-+}$ in 
$\pi^- N \to (\eta\pi^-) N$ has also been interpreted as a resonance 
[46]. Supporting evidence for the 1400 state in the same mode comes 
from ${\bar p}p \to \eta\pi^-\pi^+$ [47]. There is evidence [48]
for two isovector $0^{-+}$ states in the mass region 1.4 to 1.9 GeV;
$\pi(1600)$ and $\pi(1800)$. The quark model predicts only one. Taking the 
mass 
of the $1^{-+} \sim 1.4$ GeV, then the $0^{-+}$ is at $\sim 1.3$ GeV and the 
lightest $1^{--}$ at $\sim 1.65$ GeV, which is in the range required for the 
mixing hypothesis to work. Of course if hybrids are comparatively heavy,
that is the $\hat\rho(1600)$ is the lightest $1^{-+}$ state, and the 
$\pi(1600)$ presumably the corresponding $0^{-+}$ hybrid (or at least with
a significant hybrid component) then the vector hybrid mass $\sim 2.0$ GeV
making strong mixing with the radial and orbital excitations unlikely. 
 
Two specific models for the hadronic hybrids are the flux-tube model 
[33,40] and the constituent gluon model [49,50]. There 
are some substantial differences in their predictions for hybrid decays. 
For the isovector $1^{--}$ the flux-tube model predicts $a_1\pi$ as 
essentially the only hadronic mode, and a width of $\sim 100$ MeV. The 
constituent gluon model predicts dominant $a_1\pi$, but with significant 
$\rho(\pi\pi)_S$ and $\omega\pi$ components, and a larger width.
For the isoscalar $1^{--}$ the flux-tube model predicts $\rho\pi$ as 
essentially the only hadronic mode, with a width of $\sim 20$ MeV. The
constituent gluon model predicts dominant $\rho\pi$, a significant
$\omega(\pi\pi)_S$ component and a larger width. 

An alternative explanation could be to invoke the old concept of
multiquark states. These are defined as solutions of the multiquark
Hamiltonian with totally confined boundary conditions. In the pioneering
paper on bag-model four-quarks [51], the states were considered
with all interquark orbital momenta equal to zero. Such states easily 
decay into mesons, so that these multiquarks usually do not exist 
as relatively narrow resonances. The $q^2 \bar q^2$ states with vector
quantum numbers necessarily contain an extra unit of orbital momentum
between constituents, which could reduce the amplitude of their
"superallowed" decays. Namely, for the four-quark configurations
corresponding to the $(3 \bar 3)$ diquark-antidiquark colour representation, 
$J^{PC} = 1^{--}$ quantum numbers are achieved if orbital excitation 
$L=1$ is taken between the diquark and the antidiquark.
Extra suppresion of superallowed decay happens if the string model for
the multiquark state is adopted, in which a string with junction and
antijunction points is formed between the diquark and the antidiquark. 

In the bag model the masses of such vector states [52] lie
well above 1.7 GeV. The string model with junctions [53]
lowers the mass to 1.5 GeV giving the possibility for
$q^2 \bar q^2$ states to participate in the higher vector meson phenomena.
The detailed structure of $q^2 \bar q^2$ vector states was considerd
in [54]. The peculiar feature of the multiquark scenario is that
it is necessary to take into account three lowest states with different
total quark spins. Another interesting feature is that the lowest 
isovector state is about 200 MeV higher than the lowest isoscalar one.
Similarly to the hybrid case, selection rules for the multiquark
superallowed decay exist which forbid the decay into a pair of
$S$-wave mesons [54]. The main decay modes of $q^2 \bar q^2$ 
states are to $S$-wave plus $P$-wave mesons, and, in principle, mixing 
between $q\bar q$ and $q^2 \bar q^2$ states does the same job as mixing 
between $q\bar q$ states and hybrids. The resulting mixing scheme should
include five states, and is much more complicated than in the hybrid case.  
On the other hand it offers new opportunities, as the low-lying
four-quark $\rho(1250)$ and $\omega(1100)$ might be responsible for the
photoproduction data and the former be the ``new'' vector meson at about 
1250 MeV.

The general conclusion is that the $e^+e^-$ annihilation and 
$\tau$-decay data require the existence of a ``hidden'' vector exotics in the
isovector and isoscalar channels (assuming that the $^3P_0$ results are
qualitatively reliable). The mixing required is non-trivial, although schemes 
can be devised which are qualitatively compatible with the data [54,55]. 
The unseen physical states are ``off-stage'', in the 1.9 to 2.1 GeV mass 
region. Nonetheless, it appears difficult to achieve quantitative 
agreement with data (within the constraint of specific models) unless the 
exotics and the $1^3D_1$ states have direct electromagnetic coupling. At the
simplest level hybrids do not, but these couplings can be generated by 
relativistic 
corrections at the parton level [25] or via intermediate hadronic 
states,
for example hybrid $\to$ $a_1\pi$ $\to$ ``$\rho$'' $\to$ $e^+e^-$.
\vspace{7truemm}
\bc
{\large\bf RADIATIVE DECAYS: AN ALTERNATIVE}
\ec

Radiative decays offer several theoretical advantages. They are a much better 
probe of wave functions, and hence of models, than are hadronic decay modes
because of the direct coupling to the charges and spins of the constituents. 
This can be particularly relevant, for example, in distinguishing gluonic
excitations from conventional radial and orbital excitations as in a $1^{--}$
hybrid the $q{\bar q}$ are in a spin-singlet state which is the reverse of 
the usual $q{\bar q}$ configuration. The results of detailed calculation are 
encouraging [56]. Crucial channels can be specified and, importantly 
from the practical point of view, it is found that interesting channels
should be easily identified. The widths for radiative decays to pseudoscalar 
states are generally small, but some of those to the $1P$ states are large.
Some preliminary results are given in Table 1.
\newpage
{\small
\noindent{\bf TABLE 1.} Preliminary radiative widths in keV [56].
\bc
\begin{tabular}{|c|c|c|c|c|}
\hline
& $\Gamma(\rho_S)$ & $\Gamma(\omega_S)$ & $\Gamma(\rho_D)$ & 
$\Gamma(\omega_D)$ \\
\hline
$a_0(1300)\gamma$ & $\sim 15$ & $\sim 140$ & $\sim 110$ & $\sim 990$\\
\hline
$a_1(1260)\gamma$ & $\sim 45$ & $\sim 420$ & $\sim 80$ & $\sim 740$\\
\hline
$a_2(1320)\gamma$ & $\sim 75$ & $\sim 695$ & $\sim 12$ & $\sim 110$\\
\hline
$f_0(1300)\gamma$ & $\sim 140$ & $\sim 15$ & $\sim 990$ & $\sim 110$\\
\hline
$f_1(1285)\gamma$ & $\sim 420$ & $\sim 45$ & $\sim 740$ & $\sim 80$\\
\hline
$f_2(1270)\gamma$ & $\sim 695$ & $\sim 75$ & $\sim 110$ & $\sim 12$\\
\hline
\end{tabular}
\ec
}
\vspace{7truemm}
\noindent The larger partial widths 
should be measurable at the new high-intensity facilities. In some cases 
they may be measurable in the data from present experiments. We give two 
specific examples of quarkonia decay and a comment on hybrid radiative decay.

The $\omega\eta$ decay of the $\omega(1650)$ has been observed in the E852 
experiment [57]. If the $\omega(1650)$ is the $1D$ $q\bar q$ excitation
of the $\omega$, then the $^3P_0$ model gives the partial width for this 
decay as 13 MeV [29]. The partial width for the radiative decay $\omega(1650)
\to a_1(1250) \gamma$ is of the order of 1 MeV, that is about $8\%$ of the 
$\omega\eta$ width. The E852 experiment has several thousand events in the 
$\omega\eta$, so we may expect several hundred events in the $a_1 \gamma$ 
channel. Similarly both the $\rho(1450)$ and $\rho(1700)$ are seen by the 
VES collaboration [58] in the $\rho\eta$ channel with several thousand 
events. Both these states have strong radiative decays, the $\rho(1450)$ 
to $f_2(1270)\gamma$ and the $\rho(1700)$ to $f_1(1285)\gamma$ both of the 
order of 1 MeV. Assuming that the $\rho(1450)$ and the $\rho(1700)$ are
respectively the $2S$ and the $1D$ excitations of the $\rho$, then
the $^3P_0$ model gives the partial widths for the $\rho\eta$ decays of the 
$\rho(1450)$ and $\rho(1700)$ as 23 MeV and 25 MeV respectively [29], so the 
radiative decays should again be present at the level of a few 
hundred events.

The $\pi(1800)$ is interesting as it could be a conventional $\pi(2S)$ or 
$\pi(3S)$, the latter being the more natural if the $\pi(1300)$ is
interpreted as the $\pi(2S)$ as in the conventional quark model, or it
could be a $\pi_g$ hybrid. The $\omega(1420)$ and $\omega(1650)$ could be
conventional $2S$ and $1D$ or a hybrid $\omega_g$. The widths of the 
radiative decays $\pi(1800) \to \omega(1420)$ or $\omega(1650)$ depend 
sensitively on which of these configurations the mesons are in, and
potentially can discriminate among them. For example, if the $\pi(1800)$
is $2S$ then the width of the radiative decay to $\omega(1420)$ will be
large, 0.5 to 1.0 MeV. If it is $3S$ then the radiative decay will be 
strongly suppressed because of the orthogonality of the wave functions,
unless the $\omega(1420)$ is $3S$, which is highly unlikely. Equally, if
the $\pi(1800)$ is a hybrid, then the width of the radiative decay to the 
hybrid $\omega_g$ will again be large. Thus if a radiative width of $\sim 1$ 
MeV is found then the two states must be siblings, which would be most natural
for hybrids.

\newpage
\bc
{\large\bf SUMMARY}
\ec
Despite 15 years of work we do not yet understand the light-quark vectors.
The data raise tantalising questions which go to the heart of 
nonperturbative QCD:

$\bullet$ How many light-quark vector mesons are there?

$\bullet$ Are there exotic states hiding in there?

$\bullet$ What are the masses, widths, decay channels?

$\bullet$ Does the $^3P_0$ model fail? 

$\bullet$ If there are exotics, where are the corresponding states in the 
strange and charm sectors?

The $e^+e^-$ partial widths and the radiative decay widths of the light-quark
vector mesons provide a particularly sensitive probe of wave functions, and
the nature of the states involved. The hadronic channels test our 
understanding of decay mechanisms for radial and orbital excitations, and for
exotic states if they are present.

We are in the intriguing position of having sufficient information to realise
that the light-quark vector mesons present an exciting challenge, but have
insufficient information to solve it. Whatever the answers, new physics is 
guaranteed!
\vspace{7truemm}
\bc
{\large\bf ACKNOWLEDGEMENTS}
\ec
This work was supported in part by grant INTAS-RFBR 97-232 
\vspace{7truemm}
\bc
{\large\bf REFERENCES}
\ec
{\footnotesize
1. Erkal, C. and Olsson, M. G., {\it Z.Phys.} {\bf C31}, 615 (1986)\\
2. Donnachie, A. and Mirzaie, H., {\it Z.Phys.} {\bf C33}, 407 (1987)\\
3. Particle Data Group, {\it European Physical Journal} {\bf C15}, 1 (2000)\\
4. Achasov, M. (SND Collaboration), ``{\it Review of experimental results 
from SND}''\\
5. Pick, B. (Crystal Barrel Collaboration), ``{\it Higher vector meson 
states}''\\
6. Bartalucci, S. et al., {\it Nuovo Cimento} {\bf 49A}, 207, (1979)\\
7. Ballam, J. et al., {\it Nucl.Phys.} {\bf B76}, 375 (1974)\\
8. Atkinson, M. et al., {\it Nucl.Phys.} {\bf B243}, 1 (1984)\\
9. Brau, J. E. et al., {\it Phys.Rev.} {\bf D37}, 2379 (1988)\\
10. H1 Collaboration, Paper submitted to {\it XX International Symposium on
Lepton and Photon Interactions}, Rome, 2001\\
11. Gribov, V. N., {\it Sov.J.Nucl.Phys.} {\bf 5}, 138 (1967)\\
12. Morrison, D. R. O., {\it Phys.Rev.} {\bf 165}, 1699 (1968)\\
13. Barkov, L. M. et al., {\it Nucl.Phys.} {\bf B256}, 365 (1985)\\
Bisello, D. et al., {\it Phys.Lett.} {\bf B220}, 321 (1989)\\
14. Dolinsky, S. I. et al., {\it Phys.Lett.} {\bf B174}, 453 (1986)\\
Bisello, D. et al., {\it Proc. XXV ICHEP}, edited by K.K. Phua and
Y. Yamaguchi, World Scientific, Singapore, 1992\\
15. R. Barate et al (ALEPH Collaboration), {\it Z.Phys.} {\bf C76}, 15 (1997)\\
Perera, L. P. (CLEO Collaboration), {\it Proc. Hadron'97}, edited by S-U Chung
and H. J. Willutski, American Institute of Physics, New York, 1998, 595\\
16. Albrecht, H. et al. (ARGUS Collaboration), {\it Phys.Lett.} 
{\bf B185}, 223 (1987)\\
17. Dolinsky, S. I. et al., {\it Phys.Rep.} {\bf 202}, 99 (1991)\\
Stanco, L. (DM2 Collaboration), {\it Proc. Hadron'91}, edited by Y. Oneda
and D. Peaslee, World Scientific, Singapore, 1992, 84\\
18. Bacci, C. et al., {\it Nucl.Phys.} {\bf B184}, 31 (1981)\\
Cosme, G. et al., {\it Nucl.Phys.} {\bf B152}, 215 (1979)\\
19. Clegg, A. B. and Donnachie, A., {\it  Z.Phys.} {\bf C62}, 455 (1994)\\
Donnachie, A. and Clegg, A. B., {\it Phys.Rev.} {\bf D51}, 4979 (1995)\\
20. Achasov, N. N. and Kozhevnikov, A. A., {\it Phys.Rev.} {\bf D55}, 
2663 (1997)\\
21. Antonelli, M. et al., {\it Phys.Lett.} {\bf B212}, 133 (1988)\\
22. Fukui, S. et al., {\it Phys.Lett.} {\bf B202}, 133(1988)\\
23. Abele, A. et al. (Crystal Barrel Collaboration), {\it Phys.Lett.} 
{\bf B391}, 191 (1997)\\
24. Antonelli, M. et al., {\it Z.Phys.} {\bf C56}, 15 (1992)\\
25. Godfrey, S. and Isgur, N.,  {\it Phys.Rev.} {\bf D32}, 189 (1985)\\
26. Busetto, G. and Oliver, L., {\it Z.Phys.} {\bf C20}, 247 (1983)\\
Geiger, P. and Swanson, E. S., {\it Phys.Rev.} {\bf D50}, 6855 (1994)\\
Blundell, H. G. and Godfrey, S., {\it Phys.Rev.} {\bf D53}, 3700 (1996)\\
27. Kokoski, R. and Isgur, N., {\it Phys.Rev.} {\bf D35}, 907 (1987)\\
28. Ackleh,  E. S., Barnes, T. and Swanson, E. S., {\it Phys.Rev.} 
{\bf D54}, 6811 (1996)\\
29. Barnes, T., Close, F. E., Page, P. R. and Swanson, E. S., {\it Phys.Rev.}
{\bf D55}, 415 (1997)\\
30. Akhmetshin, R. R. et al., {\it Phys.Lett.} {\bf B466}, 392 (1999)\\
31. Anderson, S. et al., {\it Phys.Rev.} {\bf D61}, 112002 (2000)\\
32. Donnachie, A. and Kalashnikova, Yu. S., {\it Z.Phys.} {\bf C59}, 
621 (1993)\\
33. Close, F. E. and Page, P. R., {\it Phys.Rev.} {\bf D56}, 1584 (1997)\\
34. Lacock, P., Michael, C., Boyle, P. and Rowland, P., {\it Phys.Lett.} 
{\bf B401}, 308 (1997)\\
35. Bernard, C. et al., {\it Phys.Rev.} {\bf D56}, (1997) 7039 and 
hep-lat/9809087\\
36. Lacock, P. and Schilling, K., hep-lat/9809022\\
37. McNeile, C., hep-lat/9904013\\
38. Barnes, T. and Close, F. E., {\it  Phys.Lett.} {\bf 116B}, 365 (1982); 
{\it ibid} {\bf 123B}, 89 (1983)\\
Barnes, T., Close, F. E. and de Viron, F., {\it Nucl.Phys.} {\bf B224}, 
241 (1983)\\
39. Chanowitz, M. and Sharpe,S., {\it Nucl.Phys.} {\bf B222}, 211 (1983)\\
40. Isgur, N. and Paton, J. E., {\it Phys.Lett.} {\bf 124B}, 247 (1983)\\
Isgur, N., Kokoski, R. and Paton, J. E., {\it Phys.Rev.Lett.} {\bf 54}, 
869 (1985)\\
Isgur, N. and Paton, J. E., {\it Phys.Rev.} {\bf D31}, 2910 (1985)\\
Barnes, T., Close, F.E. and Swanson, E. S., {\it Phys.Rev.} {\bf D52}, 
5242 (1995)\\
41. Kalashnikova, Yu. S. and Yufryakov, Yu. B., {\it Phys. Lett.} {\bf B359}, 
175 (1995)\\
Kalashnikova, Yu. S. and Yufryakov, Yu. B., {\it Phys. At. Nucl.} 
{\bf 60}, 307 (1997)\\
42. Balitsky, I. I., Dyakonov, D. I. and Yung, A.V., {\it Z.Phys.} {\bf C33}, 
265 (1986)\\
43. Latorre, J. I., Pascual, P. and Narison, S., {\it Z.Phys.} {\bf C34}, 
347 (1987)\\
44. Weygand, D. P. (E852 Collaboration), {\it Proc. HADRON'97}, edited by 
S-U Chung and H. J. Willutski, American Institute of Physics, New York, 
1998, 313\\
45. Gouz, Yu. P. (VES Collaboration), {\it Proc. XXVI ICHEP} edited by 
J. R. Sanford, 572\\
46. Thompson, D. R. et al. (E852 Collaboration), {\it Phys.Rev.Lett.} 
{\bf 79}, 1630 (1997)\\
47. Abele, A. et al. (Crystal Barrel Collaboration), {\it  Phys.Lett.} 
{\bf B423}, 175 (1998)\\
48. Zaitsev, A. (VES Collaboration), {\it Proc. Hadron'97}, edited by 
S-U Chung and H. J. Willutski, American Institute of Physics, New York, 
(1998), 461\\
Amelin, D. V. (VES Collaboration), {\it ibid} 770\\
49. Le Yaounac, A., Oliver, L., P\`ene, O., Raynal, J. C. and Ono, S., 
{\it Z.Phys.} {\bf C28}, 309 (1985)\\
Iddir, F., Le Yaouanc, A., Oliver, L., P\`ene, O, and Raynal,J. C.,
{\it Phys.Lett.} {\bf B205}, 564 (1988)\\
50. Kalashnikova, Yu. S., {\it Z.Phys.} {\bf C62}, 323 (1994)\\
51. Jaffe, R. L., {\it Phys.Rev.} {\bf D15}, 267 (1977)\\
52. Mulders, P.J., Aerts, A.T. and de Swart, J. J., {\it Phys.Rev.} {\bf D21}, 
1370 (1980)\\
53. Badalyan, A. M., LNF-91/017(R) (1991)\\
54. Donnachie, A. and Kalashnikova, Yu. S., {\it Z.Phys.} {\bf C59}, 
621 (1993)\\
55. Donnachie, A. and Kalashnikova, Yu. S., {\it Phys.Rev.} {\bf D60}, 
114011 (1999)\\
56. Close, F.E.,  Donnachie, A. and Kalashnikova, Yu. S., in preparation\\
57. Eugenio, P. et al. (E852 Collaboration), {\it Phys.Lett.} {\bf B497} 
190 (2001)\\
58. D.V. Amelin et al. (VES Collaboration), {\it Nucl.Phys.} {\bf A668}, 
83 (2000)\\
}

\end{document}